\documentstyle[12pt,a4,cite,epsfig]{article}

\newcommand{\be}{\begin{equation}}
\newcommand{\ee}{\end{equation}}
\newcommand{\ba}{\begin{eqnarray}}
\newcommand{\ea}{\end{eqnarray}}

\renewcommand{\mathrm}[1]{{\rm #1}}

\begin{document}
\begin{titlepage}
\begin{flushright}
{CAFPE-92/08}\\
{FTUV-08-0130}\\
{UG-FT-222/08}\\
\end{flushright}
\vspace{2cm}
\begin{center}
{\large\bf   The $\sigma \to \gamma \gamma$ Width from the 
         Nucleon Electromagnetic Polarizabilities}\\
\vfill{\bf Jos\'e Bernab\'eu$^a$ and  Joaquim Prades$^b$}\\[0.5cm]
$^a$ Departament de F\'{\i}sica Te\`orica and IFIC, 
Universitat de Val\`encia - CSIC, E-46100 Burjassot, Val\`encia,
 Spain\\[0.5cm]
$^b$ CAFPE and Departamento de F\'{\i}sica Te\'orica
y del Cosmos, Universidad de Granada, Campus de Fuente Nueva, 
E-18002 Granada, Spain\\
\end{center}
\vfill
\begin{abstract}
\noindent
The  lightest QCD resonance, the $\sigma$, 
has been recently fixed in the $\pi\pi$ scattering amplitude.
 The nature of this state remains
nowadays one of the most intriguing and difficult issues
  in particle physics.
Its coupling to photons is crucial for  discriminating  its structure.
We propose a new method that fixes this  coupling 
using only available precise experimental data
on the proton electromagnetic polarizabilities 
 together with analyticity and unitarity. Taking into account
the uncertainties in the analysis and in the parameter
values, our result is
$\Gamma_{\rm pole}(\sigma \to \gamma \gamma)= (1.2\pm0.4)$ KeV.
\end{abstract}
Revised May 2008
\end{titlepage}

\noindent
     The lowest resonance in  the QCD spectrum 
has the quantum numbers of the vacuum and is
usually called the $\sigma$.  
The mass and width of this state has been recently 
fixed  with a precision of just tens of MeV in \cite{CGL06}
using an analytic continuation into the complex
energy plane of the isopin I = 0 and angular momentum 
J = 0 $\pi \pi$ partial wave scattering amplitude.
On the first Riemann sheet of the 
energy plane, the $S$-matrix  has a zero
at $E = [(441^{+16}_{-8}) -  i \, (272^{+ 9}_{-12})]$ MeV, 
which reflects the $\sigma$ pole on the second sheet at the
same position. This is also a zero at $E^*$
in the inverse of the $\pi\pi$ partial wave
 $S$-matrix  $S =1 +   2 i \, \beta(t)  \, T(t)$ 
on  the first Riemann sheet.  Here, 
\be
T(t) = \frac{1}{\beta(t) \cot(\delta(t)) +  i \beta(t)} 
\label{T}
\ee
where $\delta(t)$ is the scalar-isoscalar $\pi\pi$ phase-shift,
 $\beta(t)=\sqrt{1-4 m_{\pi}^2/t}$ and $t=E^2$.
This  result has been confirmed in Ref. \cite{GPY07} 
with the position of the $\sigma$ pole at 
$E = [(484\pm17) - i \, (255 \pm 10)]$ MeV. 
The relevance of these results has to be emphasized in view of the 
special r\^ole played by the $\sigma$ in the QCD dynamics and 
in the QCD non-perturbative vacuum structure.

Although the  pole-dominance of the $\sigma$ in  
the scalar-isoscalar $\pi\pi$ amplitude is apparent in a wide energy region
around its position, 
its existence is somewhat masked by the effects of its large width. 
     For a narrow resonance, there is an observable connection between
the phase dependence of the physical amplitude on the
real axis and the one  in the complex plane, as one crosses the pole
position. This connection is, however, lost 
in the case of the $\sigma$ with such
a large width: one does not observe either 
a rapid variation of the amplitude phase \cite{PEN07}
 nor a Breit-Wigner type behavior around the resonance position. 
This enormous difference in the behavior of the amplitude as 
one moves away from the real axis is what has made the $\sigma$ existence
and location so uncertain for so long.
    
Yet the important question 
about what is the nature of   the $\sigma$  remains unanswered
\cite{DIQ,TETRA,GLUE,TP01,TOR07,pipi,INST1,INST2,MNO08}.
What is its r\^ole in the chiral dynamics of QCD? Is it a 
$\overline{q}$-$q$ state? Is it a $\pi$-$\pi$ molecule? 
Is it a  $(\overline{qq})$-$(qq)$ tetraquark?
Is it a glueball state? How is it possible to distinguish these
different substructures? Two photon interactions can shed some light
on this question from the size of  the $\sigma \to 
\gamma \gamma$ width \cite{MMN07}. This is because this width
is proportional to the
square of the average electromagnetic charge of their constituents
while its absolute scale 
depends on how these constituents form the $\sigma$. 
Recently, the authors of \cite{PEN06,ORS08} have 
 calculated  the  $\gamma \gamma \to (\pi \pi)_{I=0,2}$ amplitudes 
using  twice-subtracted dispersion relations, 
in order to weigh the low energy region in the dispersive
integrand. Their results take into account the now well
known $\pi\pi$ final state interactions which contain the 
$\sigma$  pole in  the scalar-isoscalar contribution. 
For the width of the $\sigma$ into two-photons, 
they obtain  $(4.09 \pm 0.29)$ KeV in \cite{PEN06} and   
$(1.68 \pm 0.15)$ KeV  in the improved approach of  \cite{ORS08}.
Although the approach and methodology \cite{DIS} are very
similar in these two calculations, there is an
apparent discrepancy. Its origin   is discussed in 
\cite{ORS08}. The different input used for the dispersive 
calculation of the  production amplitudes 
of $\gamma \gamma \to \pi \pi$  
 and  the use of different values for the position
of the $\sigma$ pole on the second Riemann sheet 
$t_\sigma$ and its coupling  to two pions  $g_{\sigma\pi\pi}$ 
are  equally responsible.  Notice that although these last two
 inputs are not required in the dispersive calculation, 
the $\sigma \to \gamma \gamma$ width obtained in 
\cite{PEN06,ORS08} depends critically on them \cite{ORS08}.

The experimental results on the $\gamma \gamma \to \pi \pi$
 process are scarce and, in order to extract information
on the $\sigma$, unfortunately theoretically contaminated 
 by the Born term in the charged pion channel and by 
the isospin I = 2 amplitude in all cases,
interfering with the I = 0 amplitude in the cross section \cite{PEN07}.
     The purpose of this paper is to point out that the 
coupling $g_{\sigma\gamma\gamma}$ of the $\sigma$ meson 
 found in the $\pi\pi$ scattering  amplitude
\cite{CGL06,GPY07} is a measurable quantity, 
directly obtainable from the nucleon electromagnetic polarizabilities, 
and that it can be extracted with good precision from 
existing experimental values. This differs from the analysis in 
\cite{SCH07} where the properties of the 
$\sigma$ meson of a Nambu--Jona-Lasinio model are used.
 The  argument proceeds as follows. 
     Besides the mass, electromagnetic charge and magnetic moment,
the electric $\alpha$ and magnetic $\beta$
polarizabilities structure constants 
determine the Compton scattering amplitude 
\cite{LOW,Compton}    and the differential cross section 
up to second and  third order in the energy of the photon, respectively. 
The available experiments of Compton
scattering on protons and neutrons at low energies can be
analyzed \cite{RPP06,SCH05} in terms of $\alpha$ and $\beta$, with 
the sum $\alpha + \beta$  constrained
by the sum rule obtained from the forward dispersion relation \cite{DG70}.
The results are  $\alpha^{\rm exp} = 12.0 \pm 0.6$, 
$\beta^{\rm exp} = 1.9 \mp 0.5$
for protons and  $\alpha^{\rm exp} = 11.6 \pm 1.5$, 
$\beta^{\rm exp} = 3.7 \mp 2.0$ for neutrons. 
Here and in the rest of the paper, polarizabilities  
are given in $10^{-4} \, {\rm fm}^3$ units.

     A separate theoretical determination of $\alpha$ and 
$\beta$ needs more ingredients than the ones present in the forward 
sum rule. The authors of \cite{BT74}
investigated this problem using a backward dispersion relation
for the physical spin averaged amplitude. The corresponding sum rule for
$\alpha - \beta$ contains contributions from an s-channel part and a
t-channel part. The first is related to the multipole content of the
total photo-absorption cross section, whereas the t-channel part is
related with the imaginary part of the amplitude
through a dispersion relation for $t$, as shown  in \cite{BT77}.
This imaginary part of the amplitude is
given by the processes  $\gamma \gamma \to \pi \pi$ and 
$\pi \pi \to N \overline{N}$ via a unitarity relation. The result is the
BEFT sum rule \cite{BT74,BT77},
\ba
\alpha-\beta = \hspace*{7cm}\nonumber && \\
\frac{1}{2 \pi^2} \, \int^{\infty}_{\nu_{\rm th}}
\,  \frac{{\rm d} \nu}{\nu^2} \, \sqrt{1+ 2 \frac{\nu}{M_p}}
\, \left[\sigma(\Delta \pi = {\rm yes}) - 
\sigma(\Delta \pi = {\rm no}) \right]  && \nonumber \\
+ \frac{1}{\pi^2} \, \int^{\infty}_{4 m_\pi^2}\, 
  \frac{{\rm d} t}{4 M_p^2-t} \, \frac{\beta(t)}{t^2} \, 
\left\{ \vphantom{\frac{(4 M_p^2 - t) \, (t- 4 m_\pi^2)}{16}}
\left| f_+^0(t) \right| \, \left| F_0^0(t)  \right|
\right. \hspace*{1cm}&& \nonumber \\ \left. 
- \frac{(4 M_p^2 - t) \, (t- 4 m_\pi^2)}{16}
\, \left| f_+^2(t) \right|\, \left| 
F_0^2(t) \right| \right\} \, \hspace*{1cm}&& 
\label{BEFT}
\ea
where $M_p$ is the proton mass, 
the partial wave helicity amplitudes $f_+^0(t)$ and $f_+^2(t)$ for 
$N \overline{N} \to \pi \pi$ are Frazer and Fulco's \cite{FF60} 
and the partial wave helicity
amplitudes $F_0^0(t)$ and $F_0^2(t)$ for $\gamma \gamma \to
 \pi \pi$ are defined as in \cite{BBC76}. 
     The absorptive part in the s-channel contribution is obtained from
that of the forward physical amplitude by changing the sign of
the non parity flip multipoles ($\Delta \pi$ = no). A reliable evaluation
of this  s-channel 
integrand \cite{SCH05}  gives   $(\alpha-\beta)^s = -(5.0\pm 1.0)$
 for protons and neutrons. The importance of the t-channel contribution was
already emphasized in \cite{BT77} 
and the connection of $(\alpha - \beta)^t$ to
the isoscalar s-wave $\gamma \gamma \to \pi \pi$
 amplitude  $F_0^0(t)$ pointed out. The 
``experimental" $(\alpha - \beta)^t$ is  thus $15.1 \pm 1.3$ for protons
and $12.9 \pm  2.7$ for neutrons, compatible with the isoscalar selection
imposed by the t-channel sum rule. It is remarkable that the 
products of helicity amplitudes appearing in Eq. (\ref{BEFT})
are the products of their moduli, which might take negative values 
if the phases of these amplitudes differ from 
the $\pi\pi$ phase-shift in an odd number of $\pi$'s. 
The d-wave contribution is much smaller than the s-wave one, so that
we take it to be fixed by the Born term in the crossed 
channel \cite{HN94}, this leads to
 $(\alpha - \beta)^t_2 = -1.7$. Therefore, 
the ``experimental" quantity  to be compared with the result of the
integral  term containing  $F_0^0(t)$ in  Eq. (\ref{BEFT})
 is  $(\alpha - \beta)^t_0 =  (16.8 \pm 1.3)$.
The input  $|F_0^0(t)|$  amplitude in  that integral
is what we want to fix from this ``experimental'' value. 
The Frazer-Fulco's $|f_+^0(t)|$ amplitude 
is known with sufficient
 accuracy for our purposes from \cite{BS76} and we
have assigned  a   20 \%  uncertainty 
 to  the theoretical $(\alpha-\beta)^t_0$  
determination from the uncertainty of $|f_+^0(t)|$.
Notice that the $1/t^2$ factor in the integrand of  Eq. (\ref{BEFT})  
makes the well known  low energy 
and, to a lesser extent, intermediate energy contributions, 
to be the dominant ones. 

On the physical sheet,  we  use the twice-subtracted dispersion
 relation \cite{DIS}
\be
              F_0^0(t) \, = \, L(t) - \Omega(t)  \left[ c \, t
+ \, \frac{t^2}{\pi}
\int_{4 m_\pi^2}^{\infty} \, \frac{{\rm d} t'}{t^{'2}}
\frac{L(t') {\rm Im} \, \Omega^{-1}(t')}{t'-t-i \varepsilon} \right]
\label{DEFA}                
\ee
where $c$ is a subtraction constant fixed by
chiral perturbation theory (CHPT) \cite{DIS,GL85}, 
$c= \alpha /48 \pi f_\pi^2$ with 
$\alpha \simeq 1/137$ the fine-structure constant,
$f_\pi=92.4$ MeV the pion decay constant,
 \be
\Omega(t) = {\rm exp} \left[ \frac{t}{\pi} \int_{4 m_\pi^2}^\infty
\, \frac{{\rm d} t'}{t'} \, \frac{\delta(t')}{t'-t-i \varepsilon} \right]
\ee
is the scalar-isoscalar $\pi\pi$ Omn\`es function \cite{OMN58}
which gives the correct right-hand cut contribution 
and $L(t)$ is the left-hand cut contribution.
In this way we ensure unitarity, the correct analytic structure
of $F_0^0(t)$ and that the $\sigma$ pole properties 
enter through  the scalar-isoscalar 
phase-shift $\delta(t)$ from $T(t)$ in (\ref{T}). 
Here we shall use a simple analytic
expression for $T(t)$, compatible with Roy's equations, which takes a
three parameter fit from \cite{GPY07} 
including both low energy kaon data and 
high energy data. This fit is valid
up to values of $t$ of the order of 1 GeV$^2$, which is enough in the
integrand of the polarizability sum rule in Eq. (\ref{BEFT}).
  
At the $\sigma$ pole  position
on the first Riemann sheet \cite{PEN07,PEN06,ORS08}
\be
\label{defggamma}
F_0^0(t_\sigma) \, =  \,  e^2 \, \sqrt{6} \, 
 \frac{g_{\sigma\gamma\gamma}}{g_{\sigma\pi\pi}}\, 
 \frac{1}{2 i \beta(t_\sigma)},
\ee 
where $e$ is the electron charge, 
$g_{\sigma\pi\pi}^2$ 
is the residue of the $\pi\pi$ scattering amplitude at the $\sigma$ pole
on the second Riemann sheet and $g_{\sigma\gamma\gamma} \, 
g_{\sigma \pi\pi}$ is proportional to the residue of the $\gamma \gamma
\to \pi\pi$ scalar-isoscalar scattering amplitude on the second
Riemann sheet. The proportionality factors are such that
$g_{\sigma\pi\pi}$  
and $g_{\sigma\gamma\gamma}$ agree with those  used  in 
\cite{PEN07,PEN06}. The  pole width is given by \cite{PEN07,PEN06}  
\be
\label{width}
\Gamma_{\rm pole}(\sigma \to  \gamma \gamma) = 
\frac{\alpha^2 | \beta(t_\sigma) \, 
g_{\sigma\gamma\gamma}^2|}{4 M_\sigma} \, 
\ee
that agrees, modulo normalizations,  with that of Ref. \cite{ORS08}.
This is not the observable  radiative width that would be 
associated with a possible Breit-Wigner resonance in
 the physical $\gamma\gamma \to (\pi\pi)_{I=0}$
amplitude. However, in order to discuss the structure of the 
$\sigma$, one has to move around the pole 
and $\Gamma_{\rm pole}(\sigma \to \gamma \gamma)$ 
is the appropriate one.

Due to Low's low energy theorem \cite{LOW}, the amplitude $F_0^0(t)$
is given by the Born term at low energies. 
Then, as a first approximation,
 we consider the left-hand cut contribution $L(t)$ in (\ref{DEFA})
to be the Born contribution $L_B(t)$ to the
crossed channel describing the pion Compton scattering 
$\gamma \pi \to \gamma \pi$
\be
\label{LB}
L_B(t) = e^2 \frac{1-\beta(t)^2}{ \beta(t)} 
\, \log \left( \frac{1+\beta(t)}{1-\beta(t)} \right) \, .
\ee
Inserted into the dispersion relation 
in Eq. (\ref{DEFA}),  this contribution 
leads \cite{DIS,HN94}  to a Born  amplitude $F_0^0(t){|}_B$
for the annihilation channel $\gamma \gamma \to \pi \pi$ 
dressed with  $\pi\pi$  final state interactions. Thus, 
this $F_0^0(t){|}_B$ includes the $\sigma$ and
 is compatible with unitarity and analyticity.

\begin{figure}[ht] 
\center{\includegraphics[width=2.8in,height=1.9in]{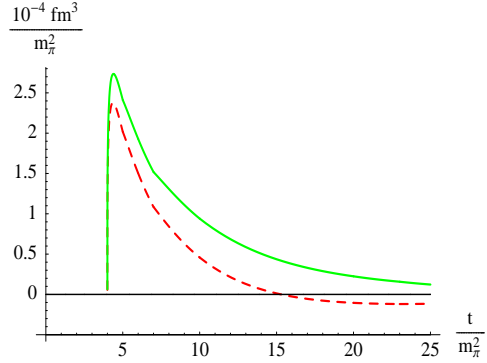}}
\caption[pilf]{\protect \small The integrand of $(\alpha-\beta)^t_0$
in (\ref{BEFT}). The broken line is when using $L(t)=L_B(t)$  in 
(\ref{DEFA}) and 
the continuous line is  when using $L(t)=L_B(t)+L_A(t)+L_V(t)$ 
in (\ref{DEFA}) as explained in the text.
\label{figure}}
\end{figure}

The evaluation of the sum rule in Eq. (\ref{BEFT}) with this
$F_0^0(t){|}_B$ results in a value 
$(\alpha - \beta)^t_0{|}_B =6.7 \pm 1.2$, 
much smaller than the ``experiment".   The quoted  uncertainty 
stems  from  the uncertainties in the input data needed 
for the sum rule in Eq. (\ref{BEFT}).
 The main reason for this small value
is the presence of a zero in the integrand of that sum rule 
 at a moderate $t$-value $t_0 \simeq 0.30 \, {\rm GeV}^2$, 
as shown in Fig. \ref{figure}.
The amplitude $F_0^0(t){|_B}$ , when analytically continued to
complex $t$, has the $\sigma$ pole in the second Riemann sheet at 
$t=t_\sigma= ([(474\pm 6)  - i \, (254\pm 4)]\,  {\rm MeV})^2$ with 
$g_{\sigma\pi\pi}= [(452\pm 4) \, +  i \,(224\pm 2) ] $ MeV
 and  on the first Riemann sheet leads to 
${g_{\sigma\gamma\gamma}/g_{\sigma\pi\pi}}{|}_B=
(0.49^{+0.03}_{-0.01})\, -\, i \, (0.37\pm0.03)$  and    
$\Gamma_{\rm pole}(\sigma \to \gamma \gamma){|}_B = 
 (2.5 \pm 0.2) $ KeV.
These results are, however, not adequate for reproducing the experimental
nucleon electromagnetic polarizabilities, as we have seen, and
the pion Compton scattering description has to go beyond
the Born approximation  $L_B(t)$, with a modification of the left-hand cut 
$L(t)$ contribution in Eq. (\ref{DEFA}).
       
At intermediate energies, this modification is due to 
 resonance  exchanges with the leading ones being  
$\gamma \pi \to a_1, \rho, \omega \to \gamma \pi$ \cite{DIS,ORS08}.
The $a_1$ exchange contribution to $L(t)$ is
\be
\label{LA}
L_A(t)= e^2 \frac{C}{32 \pi f_\pi^2} \left[ 
t + \frac{M_{a_1}^2}{\beta(t)} \log \left( \frac{1+\beta(t)+t_A/t}
{1-\beta(t)+t_A/t} \right) \right] \, 
\ee
while the $\rho$ and $\omega$ resonances exchange contribution to $L(t)$  
in nonet symmetry  ($M_\rho=M_\omega=M_V \simeq$ 782 MeV) is
\be
L_V(t)= e^2  \frac{4}{3} R_V^2   \left[ 
t - \frac{M_{V}^2}{\beta(t)} \log \left( \frac{1+\beta(t)+t_V/t}
{1-\beta(t)+t_V/t} \right) \right] \, 
\ee
 with $t_R = 2(M_R^2-m_\pi^2)$.
 The low energy limit of $L_V(t)$  goes as $t^2$  and 
we fix $R_V^2=$ 1.49 GeV$^{-2}$ by using the  well  known
$\omega \to \pi  \gamma$ decay. Though the  low energy  limit of $L_A(t)$  
goes as $t$  and  corresponds to the pion electromagnetic polarizability 
 $(\overline \alpha-\overline \beta)_{\pi^\pm}$
or equivalently to $L_9 + L_{10} = (1.4 \pm 0.3) \cdot 10^{-3}$
in CHPT \cite{BC88}, we 
 consider $L_A(t)$ as an effective contribution  for moderate higher values
of $t$ with $C$ a real constant 
to be determined phenomenologically and not connected to the pion
polarizability. This is supported by the fact that  the
 $a_1 \to \pi \gamma$ coupling is not so well known at 
intermediate energies. We fix $C$  by requiring that 
the ``experimental'' value  of $(\alpha - \beta)^t_0$ is reproduced
within 1.5 standard deviations  of the total uncertainty
when $L(t)$ in (\ref{DEFA}) is given by $L(t)=L_B(t)+L_A(t)+L_V(t)$. 
This procedure leads to $C = 0.59\pm 0.20$  
and the integrand of the sum rule is 
given in Fig. \ref{figure} as a continuous line. 
Notice that $C$ has to be positive  in order to match
the ``experimental'' value of $(\alpha-\beta)^t_0$ and 
that the zero at $t_0$ in the dressed Born amplitude has clearly 
disappeared. Moreover, in spite  of the 
fundamental dynamics of 
the $\sigma$ resonance in the t-channel polarizability sum rule, 
there is no trace of a resonant Breit-Wigner  type behavior 
when going to the physical real $t$ axis, see Fig. \ref{figure}.

The low-energy $\gamma \gamma \to \pi^0 \pi^0$ cross-sections
obtained for the  two cases  studied  above are  similar \cite{ORS08}.
The central values are compatible 
with the data for values of $t$ below 
$(450 \, \, {\rm MeV})^2$  and are above the data 
but compatible   within two standard deviations 
for larger values of $t$   up to 
$(600 \, \, {\rm MeV})^2$  and  within one  standard deviation for $t$ 
between  $(600 \, \, {\rm MeV})^2$ and $(800 \, \, {\rm MeV})^2$.

  When $F_0^0(t)$ is analytically continued to the complex plane,
at $t_\sigma$ on the first Riemann sheet one gets 
$g_{\sigma\gamma\gamma}/g_{\sigma\pi\pi}=
(0.23^{+0.05}_{-0.09}) \, - \, i \, (0.30 \pm 0.03)$
which has  a smaller absolute value when compared
with  ${g_{\sigma\gamma\gamma}/g_{\sigma\pi\pi}}{|}_B$  
 and leads to $\Gamma_{\rm pole }(\sigma \to \gamma \gamma)= 
(1.0\pm 0.3 )$ KeV.   This is the main result of this paper.
The error quoted here
is from the uncertainties in  the ``experimental'' value of 
$(\alpha-\beta)^t_0$ and the inputs of the sum rule (\ref{BEFT}) only.

In order to obtain the rest of the uncertainty, 
we  modify the $\sigma$ properties in the pion scattering
as follows. We still use the three parameter fit formula including 
low energy kaon data  and high energy  data for $\cot(\delta(t))$ 
in  \cite{GPY07}  as input in the amplitude $T(t)$
but with parameter values slightly
modified in order to reproduce the $\sigma$ pole position 
$t_\sigma= ([(441\pm 6)  - i \, (272\pm 4)]\,  {\rm MeV})^2$  
found in  \cite{CGL06}. In that case, 
we get $g_{\sigma\pi\pi}= [(480\pm 7) \, + \, i\, 
(191\pm 3)]$ MeV. 
 With this $T(t)$ and the dressed Born amplitude in (\ref{DEFA}),
one gets $(\alpha-\beta)^t_0{|}_B=6.1 \pm 1.1$,
${g_{\sigma\gamma\gamma}/g_{\sigma\pi\pi}}{|}_B=
(0.57 \pm 0.02) - i (0.41 \pm 0.03)$  and    
$\Gamma_{\rm pole}(\sigma \to \gamma \gamma){|_B}=(3.8\pm0.4)$ KeV.
The integrand of $(\alpha-\beta)^t_0$ in
(\ref{BEFT}) for this case is very similar to the broken
line of Fig. \ref{figure}.
The effective value of $C$ in (\ref{LA}) moves to $C=0.62\pm 0.20$
when fixed to reproduce the ``experimental'' value of 
$(\alpha-\beta)^t_0$ within 1.5 standard deviations of the total
uncertainty.
With this new $C$, the analytic continuation to the new $t_\sigma$
gives  $g_{\sigma\gamma\gamma}/g_{\sigma\pi\pi}=
(0.31^{+0.05}_{-0.07})\,  - \, i \, (0.32\pm0.03)$ and 
$\Gamma_{\rm pole}(\sigma \to \gamma \gamma)=(1.5\pm0.4)$ KeV. 
Again, the integrand of $(\alpha-\beta)^t_0$ in
(\ref{BEFT}) for this case is very similar to the
 continuous curve of Fig. \ref{figure}.

As final result for the electromagnetic pole width of the 
$\sigma$  found in the $\pi\pi$ scattering amplitude, we quote 
\be
\Gamma_{\rm pole} (\sigma \to \gamma \gamma) = (1.2\pm0.4)
\, \, {\rm KeV} \, 
\label{final}
\ee
 which is  the weighted average for the results of the
$\sigma \to \gamma \gamma$ width  using  
the $g_{\sigma\gamma\gamma}$ coupling in
(\ref{width}) obtained when  the $F_0^0(t)$ amplitude in  
(\ref{DEFA}) is analytically continued  to $\sigma$ pole position
$t_\sigma$ on the first Riemann sheet   (\ref{defggamma}) in two cases:
first, when  using for  $\cot(\delta(t))$ in (\ref{T}) the three-parameter
fit formula from \cite{GPY07}  including both low energy kaon 
data and high energy data;  second, when varying the
parameters of the fit for $\cot(\delta(t))$  found in \cite{GPY07}
in order to mimic the pole position found in \cite{CGL06}.
In both  cases, this $F_0^0(t)$ reproduces within 1.5 standard deviations
the ``experimental'' value of $(\alpha-\beta)^t_0$ when inserted
in the BEFT sum rule (\ref{BEFT}). 

       To conclude, we have shown that the scalar-isoscalar 
$\gamma\gamma \to \pi\pi$ amplitude $F_0^0(t)$ may be fixed 
using analyticity, unitarity and experimental information on
the nucleon electromagnetic polarizabilities. This is possible and 
direct because this component is projected out in the sum rule
(\ref{BEFT}).  When both $F_0^0(t)$  in (\ref{DEFA})
 and $T(t)$ in (\ref{T}) are
   analytically    continued to the complex plane, 
the $\sigma$ pole position and  its 
$g_{\sigma\gamma\gamma}/g_{\sigma\pi\pi}$
 and $g_{\sigma\pi\pi}$ residues become fixed.

We thank the CERN Theory Unit where this work was initiated
for warm hospitality. This work has been supported
in part by  the European Commission (EC) RTN network FLAVIAnet
Contract No. MRTN-CT-2006-035482  (J.P.), by MEC, Spain  and FEDER (EC) 
Grants  No. FPA2005-01678 (J.B.) and 
FPA2006-05294 (J.P.), Sabbatical Grant No. PR2006-0369 (J.P.),
by Junta de Andaluc\'{\i}a 
Grants No. P05-FQM 101 (J.P.), P05-FQM 467 (J.P.) and P07-FQM 03048 (J.P.)
and by the Spanish Consolider-Ingenio 2010 Programme CPAN
Grant No. CSD2007-00042.
We also would like to thank Heiri Leutwyler, 
Jos\'e A. Oller, Jos\'e Ram\'on Pel\'aez and Mike Pennington
for useful discussions and sharing unpublished results.

\end{document}